\documentclass[print]{ati}	
\usepackage{graphicx,times}
\usepackage{lmodern}
\usepackage[sort&compress, numbers,super]{natbib} 
\bibpunct[,]{[}{]}{,}{n}{}{,}			
\usepackage{url}
\usepackage[colorlinks=true,breaklinks=true, linkcolor=red,urlcolor=magenta,citecolor=blue,anchorcolor=green]{hyperref}
\usepackage{caption}
\usepackage{rotating}
\usepackage{booktabs}
\usepackage{multirow} 
\usepackage{amsmath}
\usepackage{threeparttable} 
\usepackage{CJKutf8}	
\usepackage{fancyhdr}
\usepackage{longtable}
\usepackage{pdflscape}   
\usepackage{color}
\usepackage{xcolor}
\renewcommand{\citet}[1]{\citeauthor{#1}(\citeyear{#1})\cite{#1}}	

\makeatletter
\def\@volume{}
\def\@issue{}
\def\Issue{}
\def\Year{\the\year}
\def\@pages{}
\def\Page{1}
\def\@DOI{}
\def\@citeinfo{}
\def\@date{}
\def\@copyrights{}
\def\corAuthemail{}
\def\emaillink#1{#1}
\def\corresname{}
\def\href#1#2{#2}  
\makeatother
   
\begin{document}

   \title{
   Advances and Challenges in Solar Flare Prediction: A Review
}


   \author{Mingfu Shao
      \inst{1,2}
      \ORCID{0009-0003-1447-2439}		
   \and Suo Liu
      \inst{1,2}
   \and Haiqing Xu
      \inst{1,2}
   \and Peng Jia
      \inst{3}
   \and Hui Wang
      \inst{1,2}
   \and Liyue Tong
      \inst{1}
   \and Yang Bai
      \inst{1}
   \and Chen Yang
      \inst{1,2}
   \and Yuyang Li
      \inst{2,1}
   \and Nan Li
      \inst{1,2}
   \and Jiaben Lin\correspondingAuthor{}
      \inst{1,2}
   }
\correspondent{jiaben Lin}	
\correspondentEmail{jiabenlin@bao.ac.cn}

\institute{State Key Laboratory of Solar Activity and Space Weather, NAOC, Beijing 100101, P. R. China;
          \and
              University of Chinese Academy of Sciences, Beijing 101408, P. R. China;
          \and
             College of Physics and Optoelectronics, Taiyuan University of Technology, Taiyuan 030024, P. R. China
   }
   \abstract{
Solar flares, as one of the most prominent manifestations of solar activity, have a profound impact on both the Earth's space environment and human activities. As a result, accurate solar flare prediction has emerged as a central topic in space weather research. In recent years, substantial progress has been made in the field of solar flare forecasting, driven by the rapid advancements in space observation technology and the continuous improvement of data processing capabilities. This paper presents a comprehensive review of the current state of research in this area, with a particular focus on tracing the evolution of data-driven approaches — which have progressed from early statistical learning techniques to more sophisticated machine learning and deep learning paradigms, and most recently, to the emergence of Multimodal Large Language Models (MLLMs). Furthermore, this review compares representative solar flare forecasting systems evaluated under heterogeneous settings, ranging from offline retrospective experiments to quasi-operational and operational deployments, and clarifies the relationship between reported performance metrics and real-time forecasting scenarios.
  }

   \authorrunning{ASTRONOMICAL TECHNIQUES \& INSTRUMENTS }   
   \titlerunning{Zhou A.-Y. et al.: ~Prepare a LaTeX Manuscript for ATI }  
   \maketitle
   \setcounter{page}{\Page}	
%
%
\section{Introduction}
\label{sect:intro}
Solar flares are among the most energetic manifestations of solar activity, characterized by the rapid release of magnetic energy in localized regions of the solar atmosphere, producing intense electromagnetic radiation and energetic particles that can significantly impact satellite operations, communication systems, and astronaut safety \cite{fleishman2020decay,qian2011variability}. In particular, solar flares are closely associated with coronal mass ejections (CMEs) and subsequent solar proton events (SPEs), thereby serving as potential early warning indicators of space weather disturbances \cite{youssef2012relation,garcia2016prediction,grim2024solar}. Consequently, accurate solar flare forecasting is of critical importance for safeguarding modern critical technological systems. 

\begin{figure}[!htb]
  \includegraphics[width=\linewidth]{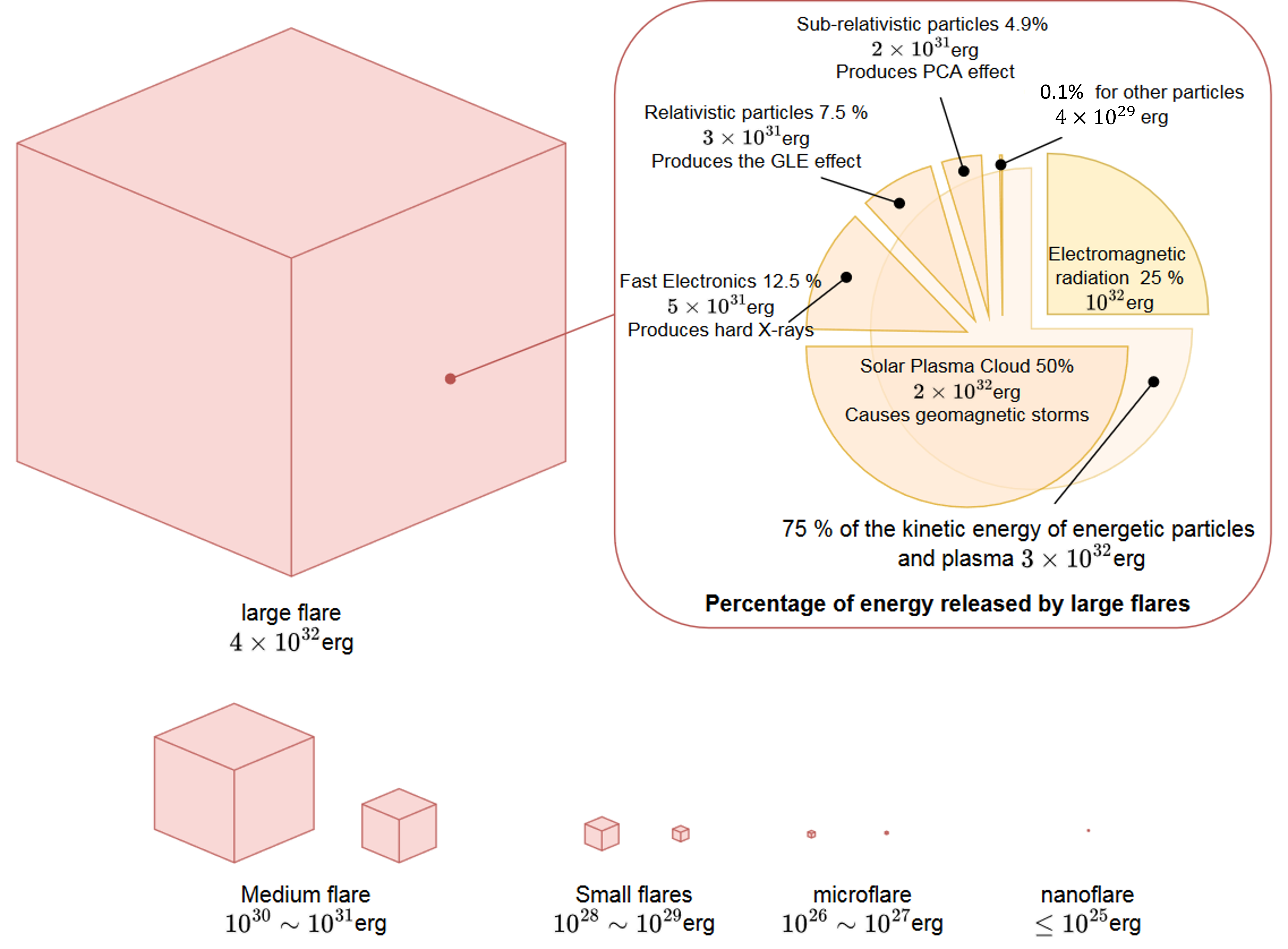}
\caption{A schematic comparison of the energy release associated with solar flares across different scales and the energy partitioning in a major event. The upper panel illustrates the distribution of different energy forms in a large flare, while the lower panel qualitatively contrasts the relative energy levels of different flare classes \cite{huang2024short}. The cubic representations are schematic and not to scale.} In particular, the boundaries between microflare and nanoflare regimes are schematic and intended only to indicate their relative positions in the flare energy hierarchy rather than precise classification thresholds.
\label{fig:flare_comparison}
\end{figure}

\begin{figure*}[!htbp]
    \vspace{2mm}\centering
    \includegraphics[width=\textwidth]{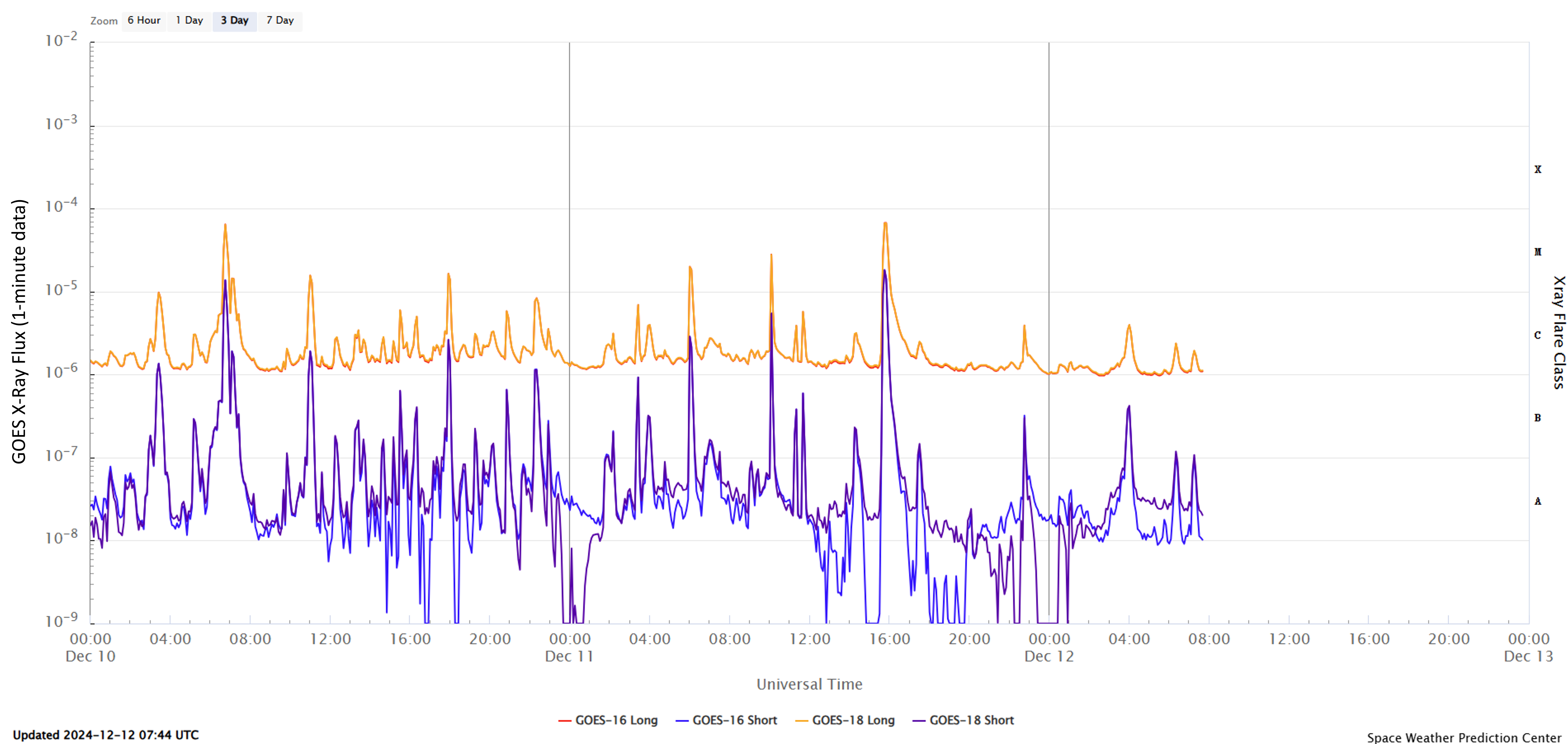} 
    \caption[]{GOES X-ray flux measured at 1-minute intervals over a three-day period. The horizontal axis denotes Coordinated Universal Time (UTC), while the vertical axis shows X-ray flux in watts per square meter (W/m$^{2}$) on a logarithmic scale. The right-hand side indicates the GOES soft X-ray flare classification scheme (A, B, C, M, X), based on the peak flux in the 1-8 Å band. The long channel (1-8 Å; 0.1-0.8 nm) of the GOES-18 and GOES-19 X-Ray Sensor (XRS) records the flux of soft X-rays, whereas the short channel (0.5-4 Å; 0.05-0.4 nm) measures the flux of hard X-rays. Data Source: NOAA/SWPC.}
    \label{fig:2}
\end{figure*}

A substantial fraction of the total energy released by a major solar flare—on the order of  $4 \times 10^{32} \, \text{erg}$—is emitted as electromagnetic radiation, constituting approximately $10^{32} \, \text{erg}$ and predominating in the visible band. The bulk of the remaining energy is converted into the kinetic energy of plasma and energetic particles \cite{ELLISON1963597, wentzel1969plasma}. At progressively smaller scales, medium-sized flares correspond to energies of $10^{30} \, $ -- $10^{31} \, \text{erg}$, and small flares to $10^{28} \, \text{erg}$ -- $10^{29} \, \text{erg}$. Further down the scale, microflares range from $10^{26} \, \text{erg}$ to $10^{27} \, \text{erg}$, while events releasing below $10^{25} \, \text{erg}$ are termed nanoflares. A schematic comparison of flare sizes and the corresponding energy distributions is illustrated in Fig.~\ref{fig:flare_comparison}. Owing to the considerable sensitivity of the Earth's ionosphere to solar soft X-ray (1-8 Å) flux variations, the soft X-ray flare class serves as a standard measure for both flare classification and quantifying associated ionospheric impacts. Solar flares are classified into five categories—A, B, C, M, and X—based on peak flux measurements in the 1-8 Å range from the Geostationary Operational Environmental Satellite (GOES), as shown in Fig.~\ref{fig:2} \cite{1994BAMS...75..757M}. Table~\ref{tab:class} presents the flux ranges for each flare class. Each class corresponds to a continuous range of soft X-ray peak fluxes measured by GOES, where the numerical value (e.g., 1.0–9.9) represents the first and second significant digits of the flux. The operational convention in solar flare prediction typically defines events of class C and above as flare events, while B-class events are generally treated as weak flares and often excluded from operational flare targets, and A-class and lower levels are regarded as quiet-Sun background emission. Furthermore, the Space Weather Prediction Center of the National Oceanic and Atmospheric Administration (SWPC/NOAA) has established the M5 level ($F_X = 5 \times 10^{-5}\ \mathrm{Watts/m^2}$) as the minimum intensity threshold for issuing official space weather alerts \cite{huang2024short}.

\begin{table}  
\renewcommand{\arraystretch}{1.3}
    \begin{minipage}[t]{0.999\linewidth}
    \caption{GOES soft X-ray classifications of solar flares}
    \label{tab:class}
    \end{minipage}
    \begin{center}
    \begin{tabular}{c c}
    \hline\noalign{\smallskip}
    \textbf{Flare class} & \textbf{Soft X-ray peak flux $(\text{Watts} \cdot \text{m}^{-2})$} \\
    \hline\noalign{\smallskip}
    A1.0--A9.9 & $(1.0$--$9.9) \times 10^{-8}$ \\
    B1.0--B9.9 & $(1.0$--$9.9) \times 10^{-7}$ \\
    C1.0--C9.9 & $(1.0$--$9.9) \times 10^{-6}$ \\
    M1.0--M9.9 & $(1.0$--$9.9) \times 10^{-5}$ \\
    X1.0 and above \textsuperscript{1} & $\geq 1.0 \times 10^{-4}$ \\
    \hline\noalign{\smallskip}
    \end{tabular}
    \end{center}
    \noindent\footnotesize
    \textsuperscript{1} The X-class flares have no upper limit; events such as X10 and X20 are also observed.
\end{table}

This article is intended as a narrative and critical review rather than a formal systematic review. The literature discussed was selected to represent major methodological developments and influential contributions to solar flare forecasting, covering physics-based approaches, machine learning and deep learning methods, and recent multimodal paradigms. The review primarily considers peer-reviewed studies published between approximately 2000 and early 2025, identified through targeted searches in major bibliographic databases such as NASA ADS and Web of Science, supplemented by cross-referencing widely cited works (a detailed summary of the search strategy and article selection process is provided in Appendix \ref{sec:appendix_a}). Given the substantial heterogeneity in datasets and evaluation protocols, the focus is on synthesizing trends and open challenges rather than providing an exhaustive, protocol-driven comparison. Studies were included based on their relevance to methodological development, peer-reviewed publication status, and demonstrated impact within the field, while works lacking sufficient methodological description or reproducibility were generally not considered.

The structure of this paper is organized as follows: Section 2 describes the data and evaluation metrics used for flare forecasting and introduces a publicly available flare prediction dataset; Section 3 provides an overview of the current research methods in the field of solar flare forecasting; Section 4 provides an applied analysis of representative solar flare forecasting systems; and finally, Section 5 summarizes the conclusions and discusses the limitations and potential directions for future research.

\section{Data and Evaluation Metrics}
\label{sect:data}
This section introduces the data used for solar flare forecasting, including a publicly available flare prediction dataset. In addition, it discusses the current limitations of existing datasets, outlines potential directions for improvement, and describes the commonly employed evaluation metrics for assessing model performance.

\subsection{Ground-based observation data}

In the era before the widespread application of satellite technology, solar flare prediction research primarily relied on observational data from ground-based observatories, which laid the foundation for early studies. For instance, Giovanelli (1939)  investigated the statistical relationship between sunspot group characteristics and solar flares using data from multiple observatories, calibrated against the standard records of the Greenwich Observatory \cite{giovanelli1939relations}. The 35-cm Solar Magnetic Field Telescope (SMFT) at the Huairou Solar Observing Station (HSOS) of the National Astronomical Observatories, Chinese Academy of Sciences, constructed in 1986, has accumulated nearly four solar cycles of vector magnetic field observations, providing the longest-running dataset of its kind worldwide and serving as the basis for numerous studies in solar flare prediction \cite{Ai1986SolarMagneticTelescope}. Similarly, the Full-disk Solar Magnetism and Activity Telescope (SMAT), established at HSOS in 2005, has also accumulated a substantial volume of observational data \cite{2007Solar}.

Ground-based observations enabled the accumulation, at an early stage, of large volumes of observational data suitable for statistical analysis, which revealed regular patterns in solar activity and laid the foundation for subsequent satellite-based studies. However, due to atmospheric seeing, weather dependence, and interruptions caused by day–night cycles, the use of ground-based data for flare prediction remains subject to certain limitations.

\subsection{Space-based Observation data}

The advent of satellite observations marked a new stage in solar physics research, greatly enhancing both the depth and breadth of solar activity data. The following section provides a detailed overview of these data sources.

\subsubsection{GOES Data}

The X-ray Flux from GOES has been a vital resource for solar flare research since its inception. This data is extensively utilized to analyze the intensity, duration, and temporal variation of solar flares across multiple solar activity cycles, particularly from cycles 23 to 25. GOES data classifies flare levels (e.g., X-class and M-class) based on peak X-ray intensity, supporting classification and regression tasks. For instance, Landa and Reuveni (2022) leveraged data from 1998 to 2019 to examine long-term trends in flare activity \cite{landa2022low}. 

Nevertheless, GOES data has several limitations. It lacks spatial observational capabilities, which makes it challenging to pinpoint the specific regions of the Sun where flares occur, thus hindering investigations into the spatial relationships between the solar magnetic field and active regions. Additionally, GOES primarily focuses on X-ray intensity and does not include other physical parameters associated with the solar magnetic field, which limits its application in more comprehensive solar activity prediction models. Despite these limitations, GOES data remains indispensable for analyzing flare intensity, duration, and temporal distribution, particularly when integrated with other observational datasets.

\subsubsection{SOHO, SDO, and ASO-S Data}

The Solar and Heliospheric Observatory (SOHO) \cite{domingo1995soho}, Solar Dynamics Observatory (SDO) \cite{pesnell2012solar}, and Advanced Space-based Solar Observatory (ASO-S) \cite{gan2019advanced} have greatly advanced solar research by providing extensive data on the solar magnetic field and active regions. The Michelson Doppler Imager (MDI) onboard SOHO, operational from 1995 to 2010, produced comprehensive line-of-sight magnetograms, while the Helioseismic and Magnetic Imager (HMI) onboard SDO, since 2010, has offered improved spatial resolution and temporal coverage, allowing more precise monitoring of magnetic field variations, which are critical for understanding magnetic reconnection and flare dynamics. More recently, the Full-disc vector MagnetoGraph (FMG) onboard ASO-S, launched by China in 2022, has provided high-resolution photospheric magnetic field, enabling detailed studies of active region evolution and flare-related magnetic processes.


Satellite missions have produced extensive datasets, greatly advancing research on solar activity and flare prediction. However, the large data volume also poses increased demands on researchers, as traditional methods are often insufficient to fully exploit these resources for reliable flare forecasting.

\subsubsection{SHARP Data Products}

The Space-weather HMI Active Region Patch (SHARP) data, derived from the SDO/HMI, provide high-level products including active region magnetograms and magnetic field parameters, such as the sum of flux near the polarity inversion line and total unsigned current helicity. These parameters are directly related to flare activity, making SHARP data particularly well-suited for developing supervised learning models and offering a valuable foundation for predicting flares and other solar phenomena. Based on the SHARP data, Abduallah et al. (2021) analyzed SHARP magnetograms along with nine associated physical parameters from May 2010 to December 2022, covering flare classifications from C- and M5-levels up to X-class events \cite{abduallah2021deepsun}.

However, SHARP data also have certain limitations. First, they only cover the period after 2010, preventing analysis of earlier solar activity. In addition, SHARP data vary in shape and size, making them unsuitable for direct use in machine learning models. They also do not perfectly correspond to solar active regions, requiring manual selection and alignment, which adds complexity to flare prediction.

\subsection{Boucheron Dataset}

In recent years, several solar flare prediction datasets have been released to support data-driven forecasting studies, based on different observational products and labeling strategies \cite{roy2025suryabenchbenchmarkdatasetadvancing,2025AMS...10553692G,boucheron2023solar,TheHMIConsortium,DVN/EBCFKM_2020}. These datasets vary in terms of data modalities, temporal coverage, and target definitions, reflecting the diversity of approaches in the flare prediction community. Among these efforts, Boucheron et al. (2023) created a publicly available solar flare prediction dataset (hereafter referred to as the "Boucheron dataset")\cite{boucheron2023solar}.The dataset is derived from line-of-sight (LoS) magnetograms of the solar photosphere, taken by the SDO/HMI. Covering the period from May 1, 2010, to December 31, 2018, the dataset consists of three distinct components: the Full Dataset, the Preprocessed Full-Resolution Dataset, and the Preprocessed Low-Resolution Dataset, which are provided to support different modeling and computational requirements. The flare labels in the dataset encompass flare intensities (e.g., "C1.0," "M1.0," "X1.0") as well as non-flare events, which are labeled as "0," with a 24-hour prediction window. In the dataset, the occurrence of a flare within the 24-hour prediction window is identified by its GOES peak time, which is adopted as a practical temporal reference for labeling purposes.

\quad a) Full Dataset (.fits format):  
This dataset contains 1,372,004 HMI active-region magnetograms extracted from 1,655 solar active regions (AR11064–AR12731). All images have a uniform spatial resolution of \(600 \times 600\) pixels. No restrictions are imposed on the heliographic latitude or longitude of the active regions. The total dataset size is approximately 537 GB.

\quad b) Preprocessed Full-Resolution Dataset (.fits format):  
This subset is derived from the Full Dataset through additional quality-control and preprocessing steps. It includes 950,047 HMI active-region magnetograms from 1,570 solar active regions, each with a spatial resolution of \(600 \times 600\) pixels. Only magnetograms whose active regions are located within a heliographic latitude and longitude range of ±60° are retained. Images containing Not-a-Number (NaN) pixels—corresponding to regions outside the visible solar disk in the FITS data—are excluded, as such cases typically occur when active regions approach the solar limb.

In addition to the magnetogram images, this dataset provides flare labels indicating whether a flare of class C1.0 or higher occurs within the subsequent 24 hours, as well as a set of 29-dimensional magnetic physical parameters for each magnetogram. These parameters consist of 7 magnetic gradient features, 13 polarity inversion line–related features, 5 wavelet-based features, and 4 magnetic flux–related features. The total dataset size is approximately 375 GB.

Specifically, researchers primarily utilize the Preprocessed Full-Resolution or Low-Resolution datasets (comprising 950,047 magnetograms from 1,570 active regions) for machine learning–based flare forecasting. The dataset is markedly imbalanced, distributed as follows: 759,465 non-flare events (label '0', 79.9\%), 161,059 C-class flares (16.9\%), 26,680 M-class flares (2.8\%), and only 2,843 X-class flares (0.3\%).

Furthermore, since a single active region contributes multiple highly correlated samples as it rotates across the solar disk, random cross-validation may lead to severe data leakage. To fundamentally resolve this, the Boucheron dataset provides pre-defined training, validation, and test sets strictly partitioned by unique active regions. An active region present in the training set will never appear in the validation or test sets. Given the massive volume of data, this strict AR-partitioned split obviates the need for traditional k-fold cross-validation while ensuring that the flare class proportions remain strictly consistent across all subsets, thereby guaranteeing a robust and unbiased evaluation.

\quad c) Preprocessed Low-Resolution Dataset (PNG format):
Derived from the preprocessed full-resolution dataset by linearly scaling the size down to \(224 \times 224\) pixels, normalizing the values to the [0,255] range, and regenerating the magnetic physical parameters. Other configurations are consistent with the full-resolution dataset, and the total size is approximately 15 GB.

Boucheron dataset provides critical support for the application of deep learning methods in solar flare prediction. However, their temporal coverage is limited to approximately eight years, falling short of encompassing even a complete solar cycle, which constrains the potential for modeling long-term evolutionary patterns. 

\subsection{Data Limitations and Improvements}

Although significant progress has been made in the availability of observational data, certain limitations still remain, which partly hinder the accuracy and generalization of predictive models.

\quad a) Class Imbalance Issue:
Solar activity datasets suffer from class imbalance, with flare events being much fewer than non-flare events. This causes machine learning models to bias predictions toward non-flare events, reducing flare prediction accuracy. While resampling and weighted loss functions help, further refinement is needed \cite{yuan2020solar,huang2018deep}. These methods adjust class distribution or assign higher importance to flare events but do not fully resolve the imbalance, requiring more research in data collection, sample generation, and algorithm development.

\quad b) Issues of Uniformity and Compatibility:

As solar flare prediction research advances, discrepancies in data preprocessing and standardization methods across studies hinder result comparison and reproducibility. For instance, Nishizuka et al. (2018, 2020, 2021) utilized data from the SDO and GOES satellites, but the preprocessing methods applied were inconsistent \citep{nishizuka2018deep,nishizuka2020reliable,nishizuka2021operational}. Similarly, Song et al. (2009) employed specific processing techniques for SOHO/MDI data, affecting the extraction of physical parameters \citep{song2009statistical}. Such methodological variations significantly influence feature representation and model performance. To enhance comparability and reproducibility, future research should aim to standardize training and testing datasets, facilitating cross-study validation.

\subsection{Evaluation Metrics:}

Evaluation metrics are employed to assess the model's performance on the test set. Solar flare prediction is generally regarded as either a classification or regression task. The primary evaluation metrics for classification tasks include the True Positive Rate (TPR)/Recall, which measures the proportion of actual positive samples correctly identified as positive; True Negative Rate (TNR), given by the ratio of correctly identified negative samples to all negatives; and Precision, calculated as the ratio of true positives to all predicted positives. In operational space weather forecasting, the False Alarm Ratio (FAR) is frequently emphasized alongside Precision. It represents the proportion of predicted flares that did not actually occur and is mathematically defined as the complement of Precision.

\begin{equation}
  TPR = Recall = \frac{TP}{TP + FN}
\end{equation}
  
\begin{equation}
  TNR = \frac{TN}{TN + FP}
\end{equation}
  
\begin{equation}
  Precision = \frac{TP}{TP + FP}
\end{equation}

\begin{equation}
  FAR = 1-Precision = \frac{FP}{TP + FP}
\end{equation}

In this context, True Positive (TP) is defined as the correct classification of an actual positive sample. False Positive (FP) denotes the incorrect classification of a negative sample as positive. False Negative (FN) refers to the incorrect classification of a positive sample as negative, and True Negative (TN) indicates the correct classification of a negative sample.

\begin{equation}
  ACC = \frac{TP + TN}{TP + FP + TN + FN}
\end{equation}
  
\begin{equation}
  TSS = TPR + TNR - 1 = \frac{TP}{TP + FN} - \frac{FP}{FP + TN}
\end{equation}
  
\begin{equation}
  F_1 = 2 \cdot \frac{Precision \cdot Recall}{Precision + Recall} = \frac{2 \cdot TP}{2 \cdot TP + FP + FN}
\end{equation}
  
\begin{equation}
  HSS = \frac{2(TP \cdot TN - FN \cdot FP)}{(TP + FN)(FN + TN) + (TP + FP)(FP + TN)}
\end{equation}

\begin{equation}
MCC = \frac{TP \times TN - FP \times FN}{\sqrt{(TP + FP)(TP + FN)(TN + FP)(TN + FN)}}
\end{equation}

When the data are balanced, Accuracy (ACC) is commonly employed as a performance metric. However, in the context of imbalanced data, such as solar flare forecasting, alternative metrics including the True Skill Score (TSS), $F_1$ score, Heidke Skill Score (HSS), and the Matthews Correlation Coefficient (MCC) are generally preferred for more robust model evaluation. Among these, the TSS is suitable for model performance comparison because it does not depend on the positive/negative event ratio of the dataset. Once the same benchmark dataset can be used, fair comparisons using other metrics will also be possible. The TSS ranges from $-1$ to $1$, where 1 indicates perfect prediction, 0 corresponds to random prediction, and negative values denote poor performance.

Similarly, the $F_1$ score—defined as the harmonic mean of Precision and Recall—is widely used in imbalanced classification tasks, as it balances false positives and false negatives. The HSS, by contrast, measures predictive skill relative to random chance, with values spanning $(-\infty, 1]$, where 1 denotes perfect classification and values near zero or negative indicate limited skill. Notably, Bloomfield et al. (2012) recommended TSS over HSS as a more appropriate metric for solar flare forecasting applications, owing to its reduced sensitivity to event imbalance \citep{bloomfield2012toward}. The MCC has also been suggested as a complementary metric for highly imbalanced flare prediction tasks, as it incorporates all entries of the confusion matrix and is less sensitive to class imbalance than accuracy-based measures \citep{Francisco_2025}. 

To systematically guide the selection of appropriate evaluation metrics based on specific data characteristics and forecasting scenarios, we summarize the recommended metrics in Table \ref{tab:metrics_guide}. As illustrated, when the dataset is relatively balanced, ACC serves as an intuitive measure of overall correctness. However, in typical solar flare datasets characterized by moderate to high class imbalance, TSS and HSS are the preferred standards for robust model benchmarking, as TSS is notably independent of the class imbalance ratio. For extremely imbalanced classification tasks, such as predicting rare X-class flares, the $F_1$ score and MCC are highly recommended because they comprehensively penalize false alarms and missed detections in the minority class. Finally, in real-world operational contexts, forecasters typically shift their focus to maximizing Recall (Probability of Detection) while strictly controlling the FAR, ensuring that major events are not missed without overwhelming the system with false alarms.

\begin{table}[htbp]
    \centering
    \caption{Guide to selecting evaluation metrics based on data features and operational contexts.}
    \label{tab:metrics_guide}
    \renewcommand{\arraystretch}{1.3}
    \begin{tabular}{c c}
        \hline
        \textbf{Data Feature} & \textbf{Recommended Metrics} \\
        \hline
        Balanced Data & ACC \\
        Moderate to High Imbalance & TSS, HSS \\
        Extreme Imbalance & MCC, $F_1$ \\
        Operational Context & Recall, FAR \\
        \hline
    \end{tabular}
\end{table}

\section{Solar Flare Forecasting Methods}
\label{sect:methods}

Short-term solar flare forecasting models typically provide predictions within 1 to 3 days prior to the occurrence of a flare. The forecast includes whether a flare will occur, intensity level, the time and location of its occurrence, and whether it will be associated with other solar activities. Solar flare prediction can be formulated either as a classification task or as a regression task, determined by the type of model outputs. Classification tasks predict discrete values, such as the likelihood of a flare occurring. Regression tasks, on the other hand, are aimed at forecasting continuous outcomes, like the peak soft X-ray flux or the total flare duration.

Based on underlying principles, solar flare forecasting methods can be further categorized into physics-based models and data-driven models. The data-driven approaches to solar flare prediction have evolved from early statistical methods to conventional machine learning techniques and, more recently, to deep neural network–based models and multimodal large language model–based frameworks \cite{huang2024short}. The performance comparison of different solar flare prediction methods is summarized in Table~\ref{tab:jw-flare-performance}. It is worth noting that conducting a strictly fair and unified comparison across existing solar flare prediction studies remains inherently challenging. In practice, most studies in the field rely on self-constructed datasets, with substantial differences in data sources, temporal coverage, train–test splitting strategies, image preprocessing procedures, and labeling definitions. As a result, establishing a common benchmark under consistent experimental settings is currently impractical. Consequently, Table~\ref{tab:jw-flare-performance} does not aim to rank methods based on their reported skill scores (e.g., TSS); rather, it compiles representative results as reported in the original publications, providing a coarse-grained and contextual overview of model performance across the literature.

\begin{sidewaystable*}
\footnotesize
\setlength{\tabcolsep}{2pt}  
\renewcommand{\arraystretch}{1.2}  
\vspace*{180mm}
  \caption{Comparison of different solar flare prediction methods within a 24-hour forecasting window. \textit{Model} denotes the backbone architecture used in each method. The parameter $t = 0.5$ indicates the probability threshold adopted for binary classification. \textit{Type} specifies the spatial coverage of active regions: AR.C refers to central active regions within predefined heliographic longitude and latitude ranges, AR.F denotes all active regions across the full solar disk, and FD represents full-disk data. The numerical labels (1, 2, 3) describe the structure of the input data, where 1 corresponds to non-spatial parameter vectors, 2 to two-dimensional spatial images, and 3 to temporal sequences of images or videos. \textit{Input} specifies the data format, with ``single'' indicating a single-frame magnetogram and ``series'' denoting a temporal sequence of magnetograms. \textit{Level} indicates the class of flares predicted by each model. \textit{Test} describes the evaluation protocol used in the original study, including the holdout method, random split-validation (RSV), and cross-validation (CV). The holdout method corresponds to offline evaluation with a fixed partition of historical data into training, validation, and testing sets; RSV allows samples to appear in both training and test sets across different splits; and CV evaluates performance by iteratively using each subset for validation. \textit{Table} indicates the corresponding table in the original paper. All evaluation metrics are extracted from the original publications. ``--'' denotes unavailable data, and ``*'' indicates values not explicitly reported in the original studies but inferred through our analysis. Caution: Performance metrics are not directly comparable across studies due to differences in datasets (temporal coverage, AR selection), preprocessing (resolution, normalization), train-test splits (random vs. AR-partitioned vs. chronological), and prediction windows. Values are reported as published and serve to illustrate methodological diversity rather than establish performance rankings.}
  \label{tab:jw-flare-performance}
  \begin{center}
  \begin{tabular}{ccccccccccccc}
      \hline\noalign{\smallskip}
      Method & Model & Type & Input & Level & Test & Table & TPR(Recall) & TNR & ACC & HSS & \textbf{TSS} \\
      \hline\noalign{\smallskip}
      Bloomfield et al. (2012) \citep{bloomfield2012toward} & Poisson Prob & AR.C 1 & Single & C+/M+/X & Holdout & T4 & 0.75/0.70/0.86 & --- & 0.71/0.83/0.88 & 0.32/0.19/0.05 & 0.46/0.54/0.74 \\
      \cmidrule[0.1pt]{2-12}
      Bobra and Couvidat (2015) \citep{bobra2015solar} & SVM & AR.C 1 & Series & M+ & RSV & T3 & 0.83 & --- & 0.92 & -0.35 & 0.76 \\
      Nishizuka et al. (2017) \citep{nishizuka2017solar} & K-NN & AR.F 1 & Single & M+/X & Holdout & T3 & --- & --- & --- & --- & 0.91/0.91 \\
      Benvenuto et al. (2018) \citep{benvenuto2018hybrid} & Hybrid & AR.F 1 & Single & C+/M+ & Holdout & T1,T2 & 0.54*/0.53* & 0.90*/0.93* & 0.91*/0.91* & 0.43*/0.46* & 0.27*/0.46* \\
      Cinto et al. (2020a) \citep{10.1093/mnras/staa1257} & Hybrid & FD 1 & Series & M+ & CV & T7 & 0.75 & 0.70 & --- & --- & 0.46 \\
      Cinto et al. (2020b) \citep{cinto2020solar} &  XGBoost  & FD 1 & Series & C+ & CV & T1 & 0.89 & 0.81 & 0.86 & --- & 0.70 \\
      Abduallah et al. (2021) \citep{abduallah2021deepsun} & Hybrid & AR.F 1 & Single & C/M/X & CV & T3 & --- & --- & --- & --- & 0.38/0.55/0.36 \\
      Ribeiro and Gradvohl (2021) \citep{2021AC} & Hybrid & AR.F 1 & Series & M+ & Holdout & T3 & 0.85 & 0.79 & --- & --- & 0.64 \\
      \cmidrule[0.1pt]{2-12}
      Nishizuka et al. (2018)\cite{nishizuka2018deep} & MLP & AR.F 1 & Single & C+/M+ & Holdout & T3 & 0.81/0.95 & 0.82*/0.85* & 0.82/0.86 & 0.53/0.26 & 0.63/0.80 \\
      Nishizuka et al. (2021)\cite{nishizuka2021operational} & MLP,t=0.5 & AR.F 1 & Single & C+/M+ & CV & T2,T3 & 0.71*/0.25 & 0.99*/0.99* & 0.99/0.99 & 0.64/0.06 & 0.70/--- \\
      Huang et al. (2018)\cite{huang2018deep} & CNN & AR.C 2 & Single & C/M/X & CV & T4 & 0.73/0.85/0.87 & 0.76/0.81/0.85 & 0.76*/0.81*/0.85* & 0.34/0.14/0.03 & 0.49/0.66/0.71 \\
      Landa et al. (2022) \citep{landa2022low} & CNN & FD 1 & Series & M/X & Holdout & T4 & 0.77/0.82 & --- & 0.85/0.93 & 0.69/0.85 & 0.69/0.85 \\
      Li et al. (2022) \cite{li2022knowledge} & CNN & AR.C 2 & Single & M+ & Holdout & T6 & 0.73 & 0.97* & 0.96 & 0.37 & 0.70 \\
      Liu et al. (2019) \cite{liu2019predicting} & LSTM & AR.C 3 & Series & C+/M+/M5+ & Holdout/CV & T3 & 0.76/0.88/0/98 & --- & 0.83/0.91/0.90 & 0.54/0.35/0.07 & 0.61/0.79/0.88 \\
      Zheng et al. (2023) \citep{zheng2023multiclass} & LSTM & AR.C 3 & Series & C+/M+/X & CV & T5 & 0.86/0.86/1.00 & --- & 0.85/0.85/0.76 & 0.69/0.60/0.18 & 0.69/0.71/0.75 \\
      Tang et al. \cite{tang2021solar} & Hybrid & AR.C 3 & Series & C+/M+ & Holdout/CV & T7 & 0.82/0.88 & --- & --- & --- & 0.64/0.72 \\
      Lv et al. (2022)\cite{inproceedings} & Hybrid & AR.C 3 & Series & C+/M+/X+ & Holdout & T2 & 0.92/0.91/0.92 & 0.96/0.86/0.99 & 0.95/0.88/0.98 & --- & --- \\
      Abduallah et al. (2023)\cite{abduallah2021deepsun} & Transformer & AR.C 1 & Series & C+/M+/M5+ & CV & T1 & 0.89/0.84/0.85 & 0.94*/0.98/0.97* & 0.92/0.93/0.96 & --- & 0.84/0.84/0.82 \\
      Li et al. (2024)\cite{li2024prediction} & Transformer & AR.C 1 & Series & M+ & CV & T6 & 0.83 & --- & 0.87 & 0.71 & 0.72 \\
      Li et al. (2025)\cite{li2025intelligent} & Transformer & AR.C 3 & Series & M+ & CV & T2 & 0.86 & 0.92 & 0.91 & 0.72 & 0.78 \\
      Francisco et al. (2025)\cite{Francisco_2025} & CNN & FD 2 & Single & C+/M+ & Holdout & T1 & 0.82/0.82 & --- & --- & 0.74/0.42 & 0.74/0.62 \\
      \cmidrule[0.1pt]{2-12}
      Roy et al. (2018)\cite{roy2025surya} & Fundation Model & FD 3 & series & M+ & Holdout & T4 & --- & --- & --- & 0.52 & 0.44 \\
      Shao et al. (2025)\cite{shao2025jwflareaccuratesolarflare} & MLLM & AR.C 3 & series & C+/M5+/X+ & Holdout & T5 & 0.84/0.97/1.00 & 0.78/0.91/0.95 & 0.79/0.91/0.95 & 0.52/0.16/0.14 & 0.61/0.88/0.95\\
      \hline\noalign{\smallskip}
  \end{tabular}
  \end{center}
\end{sidewaystable*}
\normalsize

\subsection{Physics-Based Approaches}
Physics-based approaches play a crucial role in solar flare forecasting research. These methods attempt to establish predictive models based on the physical mechanisms of the Sun, utilizing fundamental physical laws. Physical models include Magnetohydrodynamic (MHD) models and Self-organized Criticality (SOC) models.

\subsubsection{Magnetohydrodynamic Models}
MHD equations are widely used in the study of solar flares, as they enable dynamic modeling of the interactions between solar plasma and electromagnetic fields across both temporal and spatial scales \citep{piana2018flare}. Despite the success of MHD models in various fields, the physical mechanisms underlying solar flares remain incompletely understood due to the current lack of sufficient observational data and theoretical support.  Consequently, MHD models are still subject to significant limitations in predicting the occurrence and evolution of solar flares, particularly regarding high computational costs, sensitivity to boundary conditions, inadequate resolution of small-scale physics, and initialization challenges for coronal magnetic fields \citep{korsos2018applying, lin2009studies, ning2009investigation, sun2024magnetic}.

\subsubsection{Self-organized Criticality Models}
The SOC model proposed by Bak et al. (1987), explains how dynamic systems can spontaneously reach a critical state without external regulation, exhibiting 1/f noise and fractal structures, demonstrating scale invariance \citep{bak1987self}. The core idea is that small disturbances can trigger large-scale chain reactions. The SOC model focuses on self-similarity and scale invariance, rather than specific physical scales.

The SOC model has been applied to solar flare studies. Lu and Hamilton (1991) suggested that solar flares are avalanches of small magnetic reconnection events, with flare size linked to the number of reconnections \citep{lu1991avalanches}. This provides support for the solar corona’s self-organized critical state and the power-law relationship between flare frequency and magnitude. Karakatsanis and Pavlos (2008) further explored the coexistence of SOC and chaotic dynamics in solar activity \citep{karakatsanis2008soc}. Their analysis of the sunspot index revealed complex dynamical characteristics in both photospheric and sub-photospheric regions, supporting the idea of a critical state in the solar corona. 

To bridge these physical principles with modern data-driven approaches, recent studies have begun incorporating SOC concepts into machine learning architectures. For instance, Qiao et al. (2025) \citep{qiao2025flare} introduced SOC principles into a Transformer model, framing flare forecasting as a "set-prediction" task—directly predicting the number, occurrence time, and intensity of a "set" of avalanches (flares) within a given time window. They emphasized that because flares are fundamentally SOC phenomena characterized by avalanche-induced uncertainties and scale-free statistical distributions, traditional rigid binary classification (i.e., flare vs. no-flare) is suboptimal. While providing valuable insights into the complex and nonlinear characteristics of solar activity, SOC models in flare forecasting continue to undergo development.

\subsection{Data-Driven Forecasting Methods}
Data-driven forecasting methods primarily rely on observational data, extracting features through statistical analysis or machine learning algorithms to construct predictive models. 

\subsubsection{Statistical Methods}
Statistical methods predict solar flare occurrences by analyzing observational solar activity data. Giovanelli et al. (1939) first identified a statistical relationship between sunspots and solar flares, showing flare probability was positively correlated with sunspot group area, type, developmental stage, and growth rate \citep{giovanelli1939relations}. McIntosh (1990)  expanded the McIntosh sunspot classification, linking it more closely to flare occurrences than the Zurich classification \citep{mcintosh1990classification}. Wheatland et al. (2004)  introduced a Bayesian approach combining flare statistics with active region flare records to refine predictions, focusing on large flares \citep{wheatland2004bayesian}. The approach incorporated historical flare data, such as the number of small events, and included additional factors like sunspot classification.

Bloomfield et al. (2012) conducted an analysis of GOES X-ray flare observations together with the McIntosh classification during solar cycles 21 and 22 \citep{bloomfield2012toward}. They derived flare rates for each McIntosh category, estimated Poisson probabilities for various flare magnitudes, and evaluated different validation approaches.  Additionally, techniques like multiple linear regression and Fisher discriminant analysis are also commonly used in flare prediction \citep{bornmann1994flare, leka2003photospheric, gyenge2016active}. 

Traditional statistical learning methods emphasize interpretability over predictive performance and often rely on strong parametric assumptions. As a result, they exhibit limited flexibility and scalability when applied to high-dimensional or complex data, and typically depend on manually designed features and prior knowledge rather than automated representation learning.

\subsubsection{Machine Learning Methods}
In the field of solar flare prediction, machine learning methods have gradually surpassed traditional statistical learning techniques, becoming one of the dominant approaches. Several representative machine learning approaches and their applications to solar flare prediction are outlined below.

Song et al. (2009) proposed a solar flare prediction model using ordinal logistic regression \citep{song2009statistical}. By analyzing the photospheric magnetic field characteristics of solar active regions, it predicts the flare class (X, M, or C) within 24 hours. The study showed that magnetic field strength significantly affects prediction accuracy. Nishizuka et al. (2017) developed a multi-algorithm ensemble model combining Support Vector Machines (SVM), k-Nearest Neighbors (k-NN), and Extremely Randomized Trees (ERT) to predict the maximum flare class within 24 hours \citep{nishizuka2017solar}. Key features, including flare history, magnetic neutral line length, unsigned magnetic flux, and UV brightening, were found crucial for accurate predictions. Yuan et al. (2020) introduced a model integrating Principal Component Analysis (PCA) with SVM to predict solar flares \citep{yuan2020solar}. Using PCA to extract features and SVM for classification, considering sunspot activity and solar radio flux, the model achieved high accuracy and low false alarms. 

Abduallah et al. (2021) launched DeepSun, a machine learning-as-a-service (MLaaS) framework using HMI data \citep{abduallah2021deepsun}. DeepSun employs various machine learning algorithms, including Random Forest, Multilayer Perceptrons (MLP), and Extreme Learning Machines, to predict solar flares. It is the first MLaaS tool offering solar flare predictions online, combining machine learning and web services for space weather forecasting. Griffin T. Goodwin (2024) evaluated machine learning-based solar flare prediction models in simulated environments using the Space Weather Analytics for Solar Flares dataset \citep{goodwin2024investigating}. The study found that static time training windows, when sufficient data was available, could replace rolling and expanding windows, and that the solar activity cycle influenced model performance. As dataset size increased, static time windows became a more stable prediction strategy.

Machine learning methods have made notable progress in solar flare prediction by improving accuracy and automation. Nonetheless, challenges persist, particularly in feature selection. Traditional machine learning models can't automatically perform feature extraction, selection, or construction, making manual feature engineering essential for capturing complex patterns in the data, which demands substantial domain expertise and iterative adjustments. In contrast, deep learning—a major branch of machine learning—enables models to learn hierarchical representations directly from raw data, thereby reducing the need for manual feature extraction and offering greater robustness and generalization when applied to large-scale datasets.

\subsubsection{Deep Learning Methods}

Deep learning methods have been widely applied in solar flare prediction, as they can accurately capture the underlying patterns of solar active regions by learning the nonlinear mappings between raw data and predicted outcomes. Compared to traditional statistical learning and machine learning techniques, deep learning exhibits a stronger ability to automatically extract features, which enables it to excel in various prediction tasks. 

Rather than constituting a strict classification, existing deep learning approaches for solar flare prediction are more appropriately distinguished according to their dominant architectural backbones and their historical evolution. Early studies initially relied on multilayer perceptron (MLP)–based architectures \citep{1986Natur.323..533R}, and subsequently shifted toward convolutional neural network (CNN)–based \citep{1998IEEEP..86.2278L} and long short-term memory (LSTM)–based models \citep{6795963}, including hybrid architectures that combine CNNs and LSTMs. With the advent of Transformer models, more approaches have adopted Transformer-based architectures, encompassing both pure Transformer models and hybrid frameworks in which Transformers serve as the primary backbone while being integrated with other network components \citep{vaswani2023attentionneed}. More recently, this architectural evolution has further extended toward multimodal large language models (MLLMs), which leverage language-centric representations to integrate heterogeneous data modalities and enable higher-level semantic reasoning within flare forecasting frameworks.

Over a span of four years, Nishizuka et al. (2018, 2020, 2021) developed and refined an MLP-based flare forecasting model with skip connections, referred to as Deep Flare Net (DeFN) \citep{nishizuka2018deep,nishizuka2020reliable,nishizuka2021operational}. The model takes as input a set of handcrafted magnetic physical parameters derived from magnetograms and leverages multilayer perceptrons to learn nonlinear relationships among these features, thereby reducing reliance on manual decision rules while effectively modeling high-dimensional parameter spaces. Their work began with the basic DeFN model and evolved into the more reliable DeFN-R, followed by a practical application of the model for space weather forecasting. These studies significantly contributed to both the theoretical and practical aspects of solar flare prediction, offering innovative tools and insights for the field. 

In the context of solar flare forecasting, CNNs have been employed to directly analyze solar magnetograms, enabling the identification of spatial patterns such as magnetic field configurations and polarity inversion line characteristics that are closely associated with flare occurrence. 

Huang et al. (2018) introduced a deep learning-based solar flare prediction model that integrates magnetograms from SOHO/MDI and SDO/HMI with soft X-ray from GOES \citep{huang2018deep}. This model autonomously learns patterns in the data, predicting solar flares of varying intensities (C, M, X classes) over timeframes of 6, 12, 24, and 48 hours. Their study highlighted the model's stability and demonstrated that its performance was on par with other advanced prediction methods of the time. In 2022, Landa and Reuveni et al. (2022) proposed a solar flare prediction model based on a one-dimensional convolutional neural network (1D CNN) \citep{landa2022low}. Using GOES X-ray time series data spanning most of solar cycles 23 and 24, the model achieved high predictive accuracy, with particularly strong performance in forecasting M- and X-class flares. Li et al. (2022) innovatively incorporated statistical prior knowledge—such as active region (AR) area and magnetic type—into the CNN architecture, significantly improving the forecasting performance for $ \geq $ M-class solar flares \cite{li2022knowledge}. This approach validates the effectiveness of the "knowledge + data" fusion strategy and enhances the interpretability and physical consistency of knowledge-guided models.

Since solar active regions are dynamically evolving systems, their magnetic field structures and energy storage undergo changes over timescales ranging from several hours to days prior to flare onset. RNNs and their variants, such as LSTM, are well-suited for handling time-series data, enabling the capture of evolutionary trends in SHARP parameters or image sequences, thereby providing critical temporal information for flare prediction. In temporal models (such as attention-augmented LSTMs), attention weights are heavily concentrated on the critical time steps immediately preceding a flare, effectively capturing rapid pre-flare dynamics, such as sudden magnetic flux emergence or cancellation.

Jiao et al. (2020) proposed a hybrid LSTM-based deep learning framework that integrates regression and classification tasks \cite{jiao2020solar}. For the first time, this approach directly employs time series of SDO/HMI SHARP parameters to continuously predict the peak intensity of solar flares within the next 24–48 hours. Furthermore, a classification model was constructed on the basis of the regression model, yielding promising predictive performance. Liu et al. (2019)  pioneer an attention-augmented LSTM for flare forecasting, feeding 25 SHARP magnetic parameters and 15 exponentially-decayed flare-history features as 10-hour sequences into separate binary classifiers for $ \geq $ C, $ \geq $ M and $ \geq $ M5.0 events \cite{liu2019predicting}. Trained on 1,269 ARs (2010–2018) and only the top-20 features, the model attains TSS $\approx 0.88$ for $M5.0$ flares within 24 h, outperforming RF, SVM and other baselines, while demonstrating that C-class flare history is the most predictive precursor for stronger eruptions.

To simultaneously exploit spatial and temporal information, researchers have developed hybrid architectures that integrate complementary network components, typically combining CNNs for spatial feature extraction with temporal modeling modules, such as LSTM networks or other learning mechanisms. 

Tang et al. (2021) employed deep neural networks, CNN , and bidirectional LSTM to predict whether sunspot groups would produce $ \geq $ C and $ \geq $ M flare events within the next 24 or 48 hours \cite{tang2021solar}. They further demonstrated that a hybrid model, which integrates the strengths of the individual architectures, outperforms traditional statistical forecasting methods as well as any single machine learning approach. Lv et al. (2022) employ CNN and Bi-LSTM feature extraction modules to obtain high-dimensional representations that capture fused spatial–temporal features \cite{inproceedings}. Bayesian inference is then applied to predict the probability distribution of the maximum X-ray brightness within the next 24 hours. By applying thresholding, class-specific flare probabilities (C, M, and X) are derived, while Grad-CAM highlights magnetogram regions of interest, thereby enhancing both interpretability and scientific value. Zheng et al. (2023) presented a deep learning-based multi-class prediction model that combines a hybrid bidirectional long short-term memory (HBiLSTM) network with an attention mechanism \citep{zheng2023multiclass}. This model predicts no flare, C-class, M-class, and X-class flares within 24 hours. Compared to other models, including BiLSTM-Attention and CNN-based models, the HBiLSTM-Attention variant achieved the highest performance, offering superior reliability and probabilistic prediction capabilities. 

The core of Transformer models, the self-attention mechanism, enables the capture of long-range dependencies between any two elements in a sequence, making them particularly well-suited for analyzing the temporal evolution of physical parameters in solar active regions. Furthermore, visual interpretations of these mechanisms reveal that attention maps consistently highlight localized regions with strong magnetic gradients and complex topologies. Most notably, the model attention frequently localizes along highly sheared polarity inversion lines (PILs) and areas of high free magnetic energy, which aligns perfectly with the physical understanding of flare trigger sites \citep{li2025intelligent}.

Abduallah et al. (2023) introduced SolarFlareNet, a Transformer-based model designed to predict specific flare levels ($ \geq$ M5.0 , $ \geq $ M-class , $ \geq$ C-class ) in solar active regions over a 24 to 72-hour period \citep{abduallah2023operational}. Trained with flare event and SHARP magnetic field parameters from 2010 to 2022, this model outperformed existing prediction methods and has been implemented as a real-time solar flare forecasting tool. By developing multiple deep-learning architectures, Li et al. (2024) systematically compared the performance of SHARP parameters (derived from the whole active region) with HED parameters (extracted from the high-photospheric-free-energy core) for $ \geq $ M-class flare forecasting  \cite{li2024prediction}. They demonstrated that HED parameters significantly outperform SHARP parameters in both categorical and probabilistic predictions, with the Transformer model achieving the best overall performance. In 2025, Li et al. (2025) further advanced this line of research by, for the first time, introducing a single-AR magnetogram input strategy and systematically exploiting ASO-S/FMG observations for deep-learning-based flare forecasting \cite{li2025intelligent}. Their work provides a comprehensive evaluation of CNN, CNN-BiLSTM, ViT, and MViT architectures, demonstrating that the proposed MViT model achieves best performance and shows strong potential for operational deployment. 

Deep learning methods have shown significant potential in solar flare prediction, particularly in handling complex and nonlinear data, and have demonstrated improved predictive performance compared with traditional machine learning approaches. While deep learning reduces the reliance on manually engineered features, it does not eliminate the need for domain expertise, as the design of network architectures, training strategies, and hyperparameters remains a critical and specialized task, especially for large-scale and complex datasets. More recently, the emergence of large-scale models has enabled the unification of multiple data modalities, overcoming the limitations of earlier single-task solar flare prediction frameworks. Unlike conventional deep learning models that often require task-specific architecture design, large-scale models typically rely on general-purpose, pretrained architectures and can be adapted to solar flare prediction primarily through domain-specific fine-tuning. By leveraging extensive parameterization, pretrained representation capacities, and cross-modal alignment mechanisms, such models can integrate heterogeneous data sources (e.g., soft X-ray observations, magnetograms, and magnetic parameters) into a shared latent space, thereby enhancing prediction accuracy, robustness, and generalization.

\subsubsection{Multimodal Large Model Methods}

With the advent of pre-trained large language models (e.g., the GPT series \cite{ouyang2022training,achiam2023gpt} and DeepSeek-R1 \cite{guo2025deepseek}) and multimodal large language models (e.g., vision–language models \cite{bai2025qwen2,chen2024far}), such models have demonstrated strong capabilities in large-scale representation learning, cross-modal reasoning, and contextual pattern abstraction. Unlike conventional deep neural networks trained for task-specific feature extraction, these foundation models are pre-trained on diverse and heterogeneous data sources, enabling them to encode high-level semantic relationships and long-range dependencies.

In the context of solar flare forecasting, these properties offer new opportunities to integrate heterogeneous information streams, including multi-wavelength solar images, time-series magnetic field measurements, and auxiliary physical or contextual descriptions, within a unified modeling framework. Multimodal large language models can, in principle, support reasoning-oriented forecasting by jointly processing visual observations and structured or textual representations of solar activity, facilitating more flexible feature alignment and interpretability compared with purely image- or time-series–based deep learning models. As a result, MLLM-based approaches represent a promising extension beyond traditional deep learning pipelines, particularly for complex forecasting scenarios that require cross-modal information fusion and higher-level reasoning.

Roy et al. (2024) first proposed a foundation model (FM) for heliophysics, demonstrating its capability to learn universal representations from large-scale solar images and suggesting transferability to extreme event identification such as solar flares, though quantitative results were not provided \cite{roy2024ai}. Walsh et al. (2024) developed the multimodal SDO-FM, indicating that the model can recognize and process anomalies including solar flares, but emphasized that dedicated fine-tuning and validation for flare forecasting tasks require further investigation \cite{walsh2024foundation}. Roy et al. (2025) subsequently released the Surya model, which achieved pretraining on native 4096×4096 resolution solar images and, through efficient LoRA-based fine-tuning, attained a True Skill Statistic (TSS) of 0.436—significantly outperforming AlexNet and ResNet50 baselines—thereby demonstrating the robust capability of foundation models in downstream scientific applications \cite{roy2025surya}. 

Shao et al. (2025) presented JW-Flare, the first framework to adapt a multimodal large language model (MLLM) for solar flare forecasting via supervised fine-tuning \cite{shao2025jwflareaccuratesolarflare}. By leveraging a foundation model, JW-Flare effectively aligns and fuses magnetic field images with textual physical parameters into a unified semantic space. On a large-scale test dataset, the model successfully identified all X-class flares (100\% recall), achieving a TSS of approximately 0.95. To place this in an operational context, human forecasters (e.g., at NOAA/SWPC) typically achieve a TSS of 0.3–0.5 for major flares \citep{bloomfield2012toward, camporeale2025verification}. While JW-Flare demonstrates a substantial performance leap, it essentially adopts a "fail-safe" early warning strategy: it ensures zero missed extreme events at the cost of a high False Alarm Ratio (FAR =0.92). Crucially, detailed analysis shows that about 90\% of these "false alarms" corresponded to active regions that actually produced M- or C-class flares; the model correctly recognized their highly eruptive potential but merely overestimated the peak flux. Consequently, rather than producing forecasts, such MLLMs function as highly sensitive anomaly detectors that reliably flag dangerous regions to augment human forecasters in operational settings.

General spatiotemporal representations learned from large-scale unlabeled solar imagery through pretraining can be rapidly adapted to solar flare forecasting tasks via lightweight fine-tuning, effectively mitigating two major challenges in traditional approaches: the scarcity of extreme event samples and the high cost of repeated model retraining. This paradigm advances heliophysics research toward a ``one model, multiple tasks'' framework.

\section{An Applied Analysis of Representative Solar Flare Forecasting Systems}

To date, although a wide range of solar flare forecasting methods have been proposed, most studies evaluate model performance using offline experiments on historical data, such as holdout testing or cross-validation. Only a limited number of systems have been deployed as sustained operational or quasi-operational platforms that generate forecasts on unseen future data with regular automated updates and public accessibility. Table~\ref{tab:online-flare-platform} summarizes representative solar flare forecasting systems and frameworks reported in the literature, including DeepSun, DeepFlareNet, SolarFlareNet, MViT, and the NOAA/CCMC system, with emphasis on their dataset temporal coverage, evaluation protocols, and reported performance metrics. It should be noted that the reported results are obtained under heterogeneous evaluation settings, including real-time operational testing as well as retrospective experiments using historical data and cross-validation, and therefore do not uniformly represent operational performance or real-world generalization ability.

DeepSun is an online machine-learning-as-a-service (MLaaS) platform for solar flare prediction, proposed by Abduallah et al. (2021) \citep{abduallah2021deepsun}. The platform supports four-class flare prediction (B, C, M, and X) and provides both a web-based interface and an application programming interface (API) for user access. It integrates multiple machine learning algorithms and evaluates predictive performance through 10-fold cross-validation, with accuracy further enhanced by ensemble methods. 

DeepFlareNet, developed by Nishizuka et al. (2021), represents an exploratory effort to operationalize the proposed DeFN model and to evaluate its predictive performance in a real-time forecasting environment \citep{nishizuka2021operational}. The system automatically downloads near real-time data, performs preprocessing, and updates flare forecasts every six hours. It supports predictions for flares of $\geq$ C and $\geq$ M classes, with a default probability threshold of 50\%. During its operational period from January 2019 to June 2020, only four M-class flares were observed, of which one was successfully predicted, alongside 28 false alarms. Owing to the extremely limited number of M-class events, a statistically reliable skill score could not be established for $\geq$ M-class flare prediction during this period. For $\geq$ C-class flares, 38 events were recorded, with 27 successfully forecasted, corresponding to a TSS of 0.70. While the DeFN system exhibited a relatively high false alarm rate during years of low solar activity and relied on fixed model weights without online updating capabilities, its development nonetheless provided valuable experience and a benchmark reference for the future design and deployment of space weather forecasting models.

SolarFlareNet, proposed by Abduallah et al. (2023), is a Transformer-based deep learning model deployed as an operational online forecasting platform \citep{abduallah2023operational}. The system automatically updates predictions on a daily basis and provides probabilistic forecasts of flare occurrence for each active region over the next 24, 48, and 72 hours, covering $\geq$ C-class, $\geq$ M-class, and $\geq$ M5.0-class flares. Model performance was assessed in an offline setting using 10-fold cross-validation on historical data, yielding a TSS exceeding 0.83 for 24-hour forecasts and remaining above 0.7 for 48- and 72-hour horizons. These results reflect the model’s performance under retrospective evaluation and should not be interpreted as operational forecasting performance. 

\begin{table*}
  \setlength{\tabcolsep}{2pt}
  \renewcommand{\arraystretch}{1.1}
  \caption[]{Representative Solar Flare Forecasting Systems -- Summary and comparison of representative solar flare forecasting systems and platforms. Evaluation results are reported as described in the original publications; see table notes for details. EP denotes the evaluation protocol adopted in the original study; CV indicates whether cross-validation was used; RT refers to real-time operational evaluation based on unseen future data; Off-eval denotes offline evaluation conducted on historical data; SD refers to standard deviation; `*' stands for data that were not explicitly provided in the original articles but were inferred through our analysis.}
  \label{tab:online-flare-platform}
  \begin{tabular}{cccccccccccc}
  \hline\noalign{\smallskip}
  \multirow{2}{*}{\textbf{Platform}} & 
  \multirow{2}{*}{\textbf{Train}} & 
  \multirow{2}{*}{\textbf{Test}} & 
  \multirow{2}{*}{\textbf{EP}} & 
  \multirow{2}{*}{\textbf{CV}} & 
  \multirow{2}{*}{\textbf{Level}} & 
  \multirow{2}{*}{\textbf{Window}} & 
  \multicolumn{3}{c}{\textbf{Metric}} & 
  \multirow{2}{*}{\textbf{Updates\textsuperscript{5}}} \\
  \cline{8-10}
   & & & & & & & \textbf{Recall} & \textbf{Precision} & \textbf{TSS(SD)} & \\
  \hline\noalign{\smallskip}
  \multirow{4}{*}{DeepSun\citep{abduallah2021deepsun}\textsuperscript{1}} & 
  \multirow{4}{*}{2010--2016} & 
  \multirow{4}{*}{2010--2016} &
  \multirow{2}{*}{Off-eval} &
  \multirow{4}{*}{Yes} & B & 
  \multirow{4}{*}{24h} & 
  \multirow{4}{*}{---} & 
  \multirow{4}{*}{---} & 0.75 & 
  \multirow{4}{*}{False} \\
   & & & & & C & & & & 0.38 & \\
   & & & (10-folds) & & M & & & & 0.55 & \\
   & & & & & X & & & & 0.36 & \\
  \cline{6-11}
  \multirow{2}{*}{DeepFlareNet\citep{nishizuka2021operational}\textsuperscript{2}} & 
  \multirow{2}{*}{2010--2015} & 
  \multirow{2}{*}{2010--2015} &
  \multirow{1}{*}{Off-eval} &
  \multirow{2}{*}{Yes} & $\geq$ C & 
  \multirow{2}{*}{24h} & --- & --- & 0.59(0.044*) & 
  \multirow{2}{*}{6h} \\
   & & & (time-series) & & $\geq$ M & & --- & --- & 0.70(0.126*) & \\
  \cline{6-11}
  \multirow{9}{*}{SolarFlareNet\citep{abduallah2023operational}\textsuperscript{3}} & 
  \multirow{9}{*}{2010--2022} & 
  \multirow{9}{*}{2010--2022} &
  \multirow{9}{*}{Off-eval} &
  \multirow{9}{*}{Yes} &
  \multirow{3}{*}{$\geq$ C} & 24h & 0.89 & 0.95 & 0.84 & 
  \multirow{9}{*}{24h} \\
   & & & & & & 48h & 0.72 & 0.81 & 0.72(0.079) & \\
   & & & & & & 72h & 0.70 & 0.81 & 0.71(0.058) & \\
   & & & & & \multirow{3}{*}{$\geq$ M} & 24h & 0.84 & 0.85 & 0.84 & \\
   & & & & & & 48h & 0.74 & 0.82 & 0.73(0.090) & \\
   & & & (10-folds)& & & 72h & 0.71 & 0.81 & 0.71(0.095) & \\
   & & & & & \multirow{3}{*}{$\geq$ M5} & 24h & 0.85 & 0.98 & 0.82 & \\
   & & & & & & 48h & 0.74 & 0.89 & 0.74(0.112) & \\
   & & & & & & 72h & 0.72 & 0.87 & 0.73(0.108) & \\
  \cline{6-11}
  \multirow{2}{*}{DeepFlareNet\citep{nishizuka2021operational}\textsuperscript{2}} & 
  \multirow{2}{*}{2010--2015} & 
  \multirow{2}{*}{2019.01--2020.06} &
  \multirow{2}{*}{RT} &
  \multirow{2}{*}{No} & $\geq$ C & 
  \multirow{2}{*}{24h} & 0.71 & 0.60 & 0.70 & 
  \multirow{2}{*}{6h} \\
   & & & & & $\geq$ M & & 0.25 & 0.03 & --- & \\
  \cline{6-11}
  MViT\cite{li2025intelligent}  & 2010--2022 & 2023.04--2024.07 & Ensemble & No & $\geq$ M & 24h & 0.83 & 0.62 & 0.74 & False \\
  \cline{6-11}
  NASA/CCMC\cite{li2025intelligent}\textsuperscript{4} & --- & 2023.04--2024.07 & RT & No & $\geq$ M & 24h & 0.80 & 0.80 & 0.77 & 24h \\
  \hline\noalign{\smallskip}
  \end{tabular}

  \noindent\footnotesize
  \textsuperscript{1} DeepSun: \href{https://nature.njit.edu/spacesoft/DeepSun/}{https://nature.njit.edu/spacesoft/DeepSun/} \\
  \textsuperscript{2} DeepFlareNet: \href{https://defn.nict.go.jp/}{https://defn.nict.go.jp/} \\
  \textsuperscript{3} SolarFlareNet: \href{https://nature.njit.edu/solardb/index.html}{https://nature.njit.edu/solardb/index.html} (Tools $\rightarrow$ Flare Forecasting System) \\
  \textsuperscript{4} NASA/CMCC: \href{https://ccmc.gsfc.nasa.gov/scoreboards/flare/}{https://ccmc.gsfc.nasa.gov/scoreboards/flare/} \\
  \textsuperscript{5} The "Update" field indicates the system's operational update frequency. A value of "False" means the system is evaluated offline or deployed locally, rather than running as a continuously operating real-time service.
\end{table*}

In 2025, Li et al. (2025) utilized 24-hour sequences of active-region magnetograms (16 frames) and, for the first time, applied the MViT model to predict $\geq$ M-class solar flares \cite{li2025intelligent}. Using newly collected data spanning April 2023 to July 2024, they conducted a direct comparison under identical active-region samples and prediction windows against the models of the Community Coordinated Modeling Center (CCMC), a National Aeronautics and Space Administration (NASA) facility. For 24-hour $\geq$ M1.0 flare forecasts, the TSS difference between the proposed MViT model and the CCMC forecasts, evaluated over the same prediction window and active-region samples, was less than 0.03, indicating comparable performance under this specific experimental setting. Nevertheless, the MViT model remains locally deployed, without integration into real-time forecasting systems.

The CCMC constitutes a space weather modeling and validation platform under NASA. Its Flare Scoreboard module systematically aggregates 24-hour $\geq$ M-class flare probability forecasts submitted by research groups worldwide and updates them automatically at 00:00 UTC daily. Instead of providing raw observational imagery, CCMC disseminates model forecasts and associated skill scores—including TSS, recall, and precision—in a standardized format, thereby enabling fair inter-model comparison and facilitating methodological improvements. Data are freely accessible in CSV or JSON format via web interfaces or open APIs, establishing the Scoreboard as the most authoritative benchmarking and operational validation resource in the domain of solar flare prediction.

Specifically, DeepSun, SolarFlareNet, and the early-stage DeepFlareNet experiments (2010--2015) were primarily evaluated through offline retrospective experiments using historical data and cross-validation. DeepSun and SolarFlareNet both adopt 10-fold cross-validation. The early-stage DeepFlareNet experiments employ time-series cross-validation, in which training and testing sets are separated according to temporal order. While DeepSun and SolarFlareNet provide publicly accessible web-based forecasting interfaces and automated prediction pipelines, their reported skill scores are derived from offline cross-validation on historical datasets rather than continuous real-time verification on unseen future data. As such, these systems should be regarded as applied forecasting platforms with online interfaces, but their published performance metrics primarily reflect offline evaluation results. In contrast, the deployment of DeepFlareNet during 2019--2020 constitutes a genuine real-time operational experiment. Despite limitations imposed by low solar activity and insufficient statistics for $\geq$M-class flares, this effort provides an important reference for real-time solar flare forecasting.

The MViT model represents a distinct category. It is an advanced deep learning model evaluated under a controlled real-time-style comparison against CCMC forecasts, using newly collected data and identical prediction windows. However, the MViT system remains locally deployed and should be regarded as an advanced research model rather than a fully operational platform. The MVIT model employs an ensemble strategy: ten separate models are trained using 10-fold cross-validation on historical data (2010-2022), but performance is assessed on a strictly held-out future period (April 2023 - July 2024) by averaging probabilistic outputs from all ten models. This constitutes ensemble-based evaluation on unseen future data rather than traditional cross-validation, thereby approximating operational forecasting conditions. Finally, the NASA/CCMC Flare Scoreboard provides a vital platform for operational benchmarking by aggregating real-time forecasts from multiple models. It is important to note that the TSS value of 0.77 and other evaluation metrics referenced here is derived from the comparative analysis in Li et al. (2025) \cite{li2025intelligent}, where the authors aggregated the probabilistic outputs of six CCMC-participating models to compute standardized skill scores on unseen data, thereby enabling fair inter-model comparison.

Overall, this section reviews representative solar flare forecasting systems, spanning the full spectrum from offline research-oriented models to quasi-operational and fully operational forecasting platforms. The systems summarized in Table~\ref{tab:online-flare-platform} are included not because they all constitute sustained real-time services, but because they exemplify different stages of methodological maturity, deployment readiness, and evaluation paradigms within the field.

\section{Conclusion}
\label{sect:conclusion}

This review traces the evolution of forecasting methodologies from physics-based models to data-driven techniques, and summarizes the model performances within a 24-hour prediction window.

\textbf{Data perspective:} We survey space-based observations, including GOES, SOHO/MDI, SDO/HMI, and ASO-S/FMG, as well as ground-based observations such as SMFT and SMAT, focusing on magnetograms, X-ray measurements, and derived magnetic parameters. Earlier publicly available datasets (e.g., Boucheron) were limited in temporal coverage (often less than a solar cycle), characterized by severe class imbalance, and subject to inconsistent preprocessing standards. With the ongoing release of extended SDO observations spanning 2010–2025, these constraints are being progressively alleviated, opening new opportunities for cross-cycle generalization and more systematic model evaluation.

\textbf{Methodological perspective:} Physics-based approaches (e.g., MHD, SOC models) provide mechanistic interpretability but remain inadequate for meeting the timeliness and accuracy required in 24-hour operational forecasting. Statistical and traditional machine learning methods, while stable, rely heavily on handcrafted features and are constrained in their ability to capture high-dimensional nonlinear dynamics. Deep learning methods have addressed some of these challenges: CNNs extract spatial patterns from magnetograms, RNN-LSTMs capture temporal evolution, and hybrid architectures integrate spatial and temporal information. Transformer-based models, leveraging self-attention, have achieved notable advances in modeling long-range dependencies. More recently, MLLMs trained via self-supervised pretraining and adapted through lightweight fine-tuning, have demonstrated strong performance even under data-scarce conditions, highlighting their potential for unified “multi-task” prediction frameworks. Importantly, the evaluation of data-driven approaches, particularly multimodal large language models, must explicitly enforce strict temporal separation when using SDO/HMI data to prevent information leakage.

\textbf{Applied performance:} Current operational or quasi-operational services differ substantially in their input data sources, forecast targets, prediction cadences, and output formats, reflecting distinct design objectives rather than a unified evaluation framework. Most platforms primarily rely on offline-trained models and retrospective validation using historical datasets, while only a limited number of systems have been tested under sustained near-real-time forecasting conditions. As a result, reported performance metrics are often derived from controlled experimental settings and should be interpreted with caution when extrapolated to operational environments. Importantly, recent advances in long-duration SDO observations and the gradual emergence of real-time forecasting services provide a practical foundation for improving applied evaluation in the future. These developments enable more systematic testing of model robustness, stability, and reliability under realistic space weather conditions, which will be critical for bridging the gap between research-oriented prediction models and operational space weather forecasting.

To conclude, despite notable advances in solar flare forecasting, important challenges remain, and future progress is expected to focus on the following directions:
(1)  The construction of authoritative benchmark datasets spanning multiple solar cycles is essential for fair comparison and accelerated algorithmic progress. Achieving this requires carefully addressing the technical distinctions between varying observational cadences and broader data standardization approaches when integrating high spatio-temporal resolution multimodal data. For instance, a robust benchmark integrating SOHO/MDI (1996–2010, 96-min cadence, 1024×1024 resolution) and SDO/HMI (2010–present, 12-min cadence, 4096×4096 resolution) magnetograms  must apply standardized cross-calibration protocols. To maintain temporal consistency and physical integrity without introducing interpolation artifacts, SDO/HMI observations can be temporally subsampled to a common 96-minute cadence and spatially downsampled to a 1024×1024 resolution. Furthermore, adopting active-region-partitioned or chronological train/validation/test splits is crucial to prevent data leakage.
(2) The development of models with physically interpretable predictive mechanisms—such as Physics-Informed Neural Networks (PINNs), hybrid MHD-ML frameworks, or advanced attention mechanisms that explicitly highlight localized physical precursors—is of primary importance, both for uncovering novel precursors of flare activity and for fostering trust in AI-based forecasting among human forecasters. For instance, PINNs could incorporate MHD constraints (e.g., $\nabla \cdot \mathbf{B} = 0$, magnetic pressure balance) as physics-informed loss terms, ensuring that learned magnetic field reconstructions remain physically consistent while data-driven components capture complex flare precursors not explicitly modeled in traditional MHD simulations.
(3) End-to-end online learning platforms, incorporating reinforcement learning and large-model agent architectures, will be critical for enabling continuous adaptation to solar-cycle variability and for enhancing real-time forecast reliability.
(4) The exploration of multi-task large models tailored for space weather applications—capable of jointly predicting flare class, peak flux, CME association probability, and likely source regions—offers a promising pathway toward an “inference-once, multi-product” operational paradigm.

\section*{acknowledgements}

This research was supported by the National Astronomical Observatories, Chinese Academy of Sciences (No. E4TQ2101), and the National Natural Science Foundation of China (No. 12473052), with additional support from the Specialized Research Fund for the State Key Laboratory of Solar Activity and Space Weather. The authors also acknowledge the Association for Astronomy × A.I. (A³), funded by the Science and Education Integration Program of the University of Chinese Academy of Sciences, and appreciate the support provided by the School of Astronomy and Space Science, University of Chinese Academy of Sciences.

\section*{Author Contributions}
Lin conceived the ideas. Shao implemented the study and wrote the paper. Wang collected the data. All authors read and approved the final manuscript.
 
\section*{Declaration of Interests}
The authors declare no competing interests.

\appendix                  
\section{Literature Search Strategy}
\label{sec:appendix_a}

Table \ref{tab:search_strategy} details the search terms, date ranges, and the article selection process used to identify the representative studies discussed in this review.

\begin{table*}[htbp]
    \centering
    \caption{Summary of the literature search and selection process.}
    \label{tab:search_strategy}
    \renewcommand{\arraystretch}{1.5} 
    \begin{tabular}{p{0.20\textwidth} p{0.55\textwidth} p{0.20\textwidth}}
        \hline
        \textbf{Phase} & \textbf{Details / Criteria} & \textbf{Approximate Count} \\
        \hline
        \textbf{Targeted Databases} & NASA ADS, Web of Science & - \\
        
        \textbf{Date Range} & January 2000 to February 2025 & - \\
        
        \textbf{Core Search Terms} & 
        ("solar flare prediction" OR "solar flare forecasting") AND ("machine learning" OR "deep learning" OR "magnetohydrodynamics" OR "MHD" OR "large language model") & 
        Initial identification: $\sim$1,500 articles \\
        
        \textbf{Screening \& Inclusion Criteria} & 
        1. Peer-reviewed journal publications.\newline 
        2. High relevance to methodological development (e.g., transition from statistical to deep learning/MLLMs).\newline 
        3. Sufficient methodological description and dataset transparency. & 
        Excluded articles lacking clear methodological descriptions, redundant empirical studies, or those strictly focused on non-flare phenomena. \\
        
        \textbf{Final Inclusion} & Selected to represent major methodological paradigms and influential contributions. & Final included: 50 articles \\
        \hline
    \end{tabular}
\end{table*}


\clearpage

\bibliographystyle{plain}
\bibliography{ati}      

\label{lastpage}

\end{document}